# Unravelling the electrical properties of epitaxial Graphene nanoribbons

Z. Xiao & C. Durkan[a]

Nanoscience

University of Cambridge

11 JJ Thomson Avenue

Cambridge, CB3 0FF

UK

The electrical resistivity of single-layer graphene nanoribbons has been studied experimentally for ribbon widths from 16 nm – 320 nm and is shown to validate the expected quantum scattering model for conduction through confined graphene structures. The experimental findings are that the resistivity follows a more dramatic trend than that seen for metallic nanowires of similar dimensions, due to a combination of the nature of the charge carriers in this 2D material, surface scattering from the edges, band-gap related effects and shifts in the Fermi level due to edge effects. We show that the Charge Neutrality point switches polarity below a ribbon width of around 50 nm, and that at this point, the thermal coefficient of resistance is a maximum. The majority doping type therefore can be controlled by altering ribbon width below 100 nm. We also demonstrate that an alumina passivation layer has a significant effect on the mean free path of the charge carriers within the graphene, which can be probed directly via measurements of the width-dependent resistivity.

[a] : corresponding author, email cd229@cam.ac.uk

There has been much interest in the properties of graphene since its isolation in 2004 [1, 2]. This is due in part to the strength of the C-C bond which leads to an in-plane Young's modulus of order 1 TPa [3], and also to the unique electrical properties resulting from the specific band structure of this 2D material. Graphene devices suffer from a host of issues in that although the charge carriers have zero effective mass and can travel ballistically over large distances exceeding tens of microns, this is rarely seen in practice due to the influence of defects. Unlike similarly-sized metal structures, electrical transport in graphene is entirely dominated by surface effects for obvious reasons.

Given that the surface of graphene is invariably covered with contaminants from the atmosphere and chemical residues from device fabrication, this has a profound effect on its electrical properties with the result that graphene is often unintentionally doped *p*-type. As a result of this the electrical resistivity is often orders of magnitude lower than expected and the contact resistance is often rather high [4-5], and the appropriate choice of metal electrode material is critical in order to avoid a Schottky-type contact or any band-bending. On the one hand, this propensity for doping potentially makes graphene desi able as a gas sensor [6, 7] but on the other hand, it is difficult to control, and means that we often cannot realise its true potential of having ultra-high mobility ballistic charge carriers. Even in the absence of any ambient contaminants, the underlying substrate can have the effect of doping graphene, for example graphene on pristine $SiO_2$ has been shown to be *n*-type [8]. Of course, one can carry out investigations on atomically-clean, suspended devices under UHV and low-temperature conditions, but this is neither scalable nor commercially practical. Recent efforts have focused on ways of passivating the surface of graphene to mitigate against such unintentional doping, with varying degrees of success [5, 9-11]. This typically involves coating the graphene with a thin layer of oxide, commonly $Al_2O_3$ or $TiO_2$ or a nitride such as HBN. The mode of deposition of this layer also plays a role in determining the electrical characteristics of any devices thus made as this determines the nature and prominence of defects. It has been shown [12] that ALD-deposited $Al_2O_3$ has fewer defects than thermally-evaporated $Al_2O_3$ and therefore graphene devices coated with it exhibit higher resistivity due to lower unintentional doping levels. Ultimately, the presence of the defects on the top and bottom surface leads to scattering of the electrons and holes within graphene, and this gives rise to an effective mean free path, $\lambda$ that is significantly lower than the intrinsic one. This is similar to what happens in a doped semiconductor where the effect of doping is to increase the number of charge carriers which ultimately increases the conductivity, but the carriers end up with a reduced mobility and mean-free path due to the presence of the dopants.

Coupled with the fact that single-layer graphene (SLG) has no intrinsic bandgap and a relatively low current on/off ratio [13], it is clear that there are only very limited applications for this material in bulk form. However, given that electrons in graphene can have relatively long coherence lengths of up to several hundred nm [14], one can make use of quantum size effects to artificially induce a bandgap. It has been shown in a number of reports in recent years that such quantum-confined structures, known as graphene nanoribbons (GNR) have a bandgap ($\Delta E$) that scales as $\Delta E \propto \frac{1}{w}$ where $w$ is the width of the ribbon, as expected from the well-known quantum particle (in this case a massless Dirac Fermion with linear dispersion, rather than the conventional quadratic dispersion as seen in a metal or semiconductor, where bandgap scales as $\frac{1}{w^2}$) in a box. This bandgap is of order 100-200 meV for 10 nm wide GNRs, reducing as the width decreases. Studies have also shown that the effective electron mobility appears to depend on GNR width [15], decreasing as ribbons get narrower. This is the same as saying that scattering and resistivity increase as ribbon width decreases. It should be pointed out that the same behaviour has been observed in metal nanowires for decades, but has not been described in terms of mobility, and is due to a combination of surface and grain-boundary scattering.

As well as the top and bottom surfaces, the edges of graphene are also responsible for scattering. It has been shown both theoretically and experimentally [16-18] that the edge termination has a significant effect on the resistivity, with armchair edges leading to greater scattering than zig-zag edges. It was subsequently shown [19] that this is due to the fact that for the specific case of zig-zag edges with no disorder or chemical functionalisation, the electron wavefunctions are zero at the edge, so sliding electrons (i.e. those travelling along the edge) experience no scattering. However, for the case of some edge disorder, induced by either chemical modification/functionalisation or edge roughness, both of which result from the processing steps required to fabricate the GNRs, there can be significant edge scattering. The fact that we observe no difference in resistivity of GNRs fabricated at different orientations within the same graphene grain indicates that the level of edge disorder is indeed significant, and the edges are not exclusively either zigzag or armchair. Given that the Fermi wavelength of electrons in graphene is of order 1 nm, it is no surprise that overall, scattering at the edges of patterned graphene is almost fully diffuse [14]. There is also the effect of [13] reduced effective GNR width due to doping-induced charge depletion at the edges that can extend several nm into the GNR from either edge. There is continued interest in fabricating all-graphene devices, where the active (doped) regions as well as the interconnects are all fabricated using graphene. In order for this to lead to devices that can replicate the functionality of conventional CMOS devices, and in order to have tuneable bandgaps and useful on/off ratios, the scale of these devices must therefore be in the sub. 100-nanometer range, wherein edge effects will inevitably have an adverse influence on the transport.

In this article, we explore the combined effect of surface coating and ribbon width on the resistivity of graphene ribbons with or without passivation layers of $Al_2O_3$ deposited by electron beam evaporation. Previous experimental studies have concentrated on larger ribbons and not provided a detailed explanation for their observations, and theoretical studies have not been verified experimentally. Therefore, we have set out to perform a systematic experimental study of resistivity versus size, and show that a simple model for transport which takes edge and size effects into account can be used to explain our findings, laying the ground for further studies. Similar studies on metal interconnects [20, 21, 22] have shown that microstructure plays just as important a role in determining resistivity as the wire cross-section, particularly for widths/thicknesses around and below the electronic mean-free path. We anticipate that for graphene, as it has very little microstructure apart from folds and occasional grain boundaries, we will instead be sensitive to the properties of the materials in contact with the top and bottom surfaces as well as the intriguing effects due to doping/disorder at the edges, coupled with the bandgap introduced by the lateral confinement.

The problem of size and surface-related conductivity effects in electrical materials has been around for more than 80 years. It is known that the resistivity of metallic thin films increases as soon as the film thickness decreases below the effective electronic mean free path. In the mesoscopic regime where structures are large enough that discrete quantum effects are not noticeable, this effect was attributed to diffuse scattering at the film boundaries by Fuchs and Sondheimer [23, 24]. This scattering leads to the notion of an *effective* mean free path which then depends on thickness. As resistivity is inversely proportional to mean free path, the resistivity consequently increases as dimensions reduce. This model has its roots in the semi-classical Boltzmann transport equation and describes how the effective mean free path is modified in the presence of surface scattering based on geometric arguments. As a result, it should be possible to apply it to metals, semiconductors and even graphene. One must be careful when investigating electronic transport at these small lengthscales as they are comparable to the mean-free path, $\lambda$, so the transport is part ballistic, requiring analysis within the quantum regime, i.e. the Landauer-Büttiker formalism [25, 26], while also being partly diffusive.

In order to gain an understanding of the size-dependence of the electrical resistivity of graphene, we fabricated a series of single-layer graphene ribbons with widths ranging from 16 to 320 nm. The wires were prepared by a multistep process involving electron beam lithography, oxygen plasma etching and metallization, as shown in Figure 1.

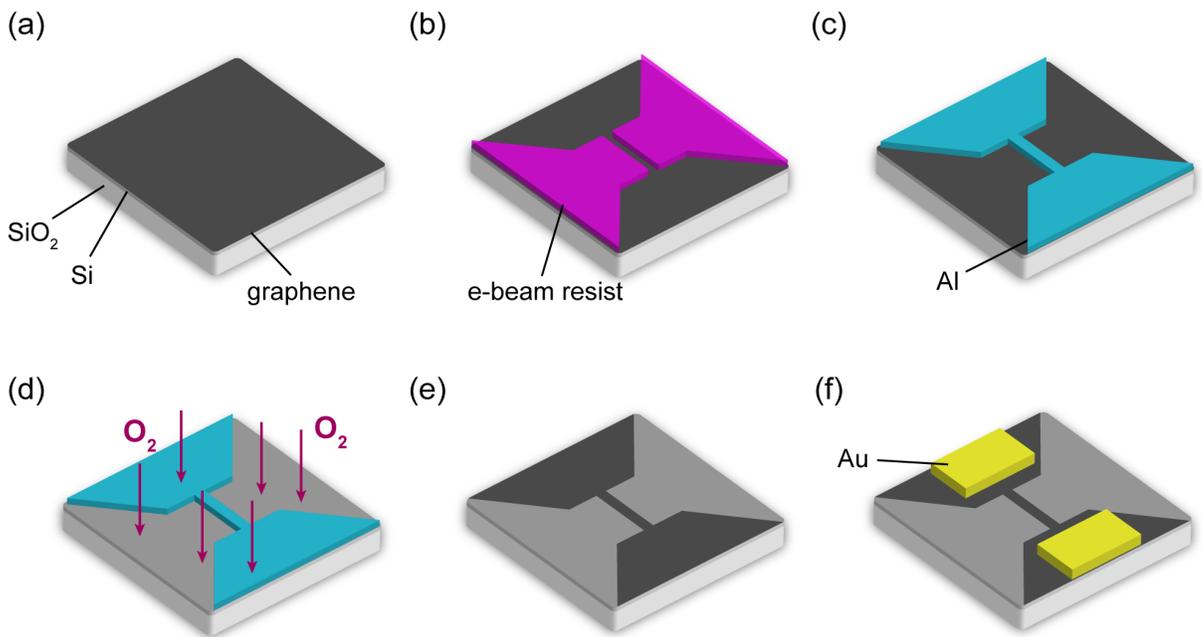

*Figure 1.* *Process flow. (a) Graphene is transferred onto 300nm thick $SiO_2$/Si; (b) e-beam resist is spin-coated, exposed and then developed leaving a template for (c) evaporation of a 7.5 nm thick layer of Al, which is oxidised to form insulating $Al_2O_3$; Sample is then exposed to an Oxygen plasma which removes the graphene everywhere apart from underneath the $Al_2O_3$ which is then (e) removed using HCl; (f) in the final step, Au/Cr electrodes are lithographically patterned on top of the graphene device.*

The monolayer graphene was grown on Cu foil (predominantly (110)) substrates by CVD (Chemical Vapor Deposition) and then transferred onto 300nm thick $SiO_2$ on p-doped Si substrates by a wet transfer method. These were then spin-coated with electron beam resist PMMA 950A2 (70nm thick) and baked at 200ºC for 2mins. Graphene ribbons of different width and length, *w & L*, respectively were created using a Crestec CABL-9000 High Resolution Electron Beam Lithography System using 100pA beam current and 50kV acceleration voltage. Development was performed in 3:7 water: isopropanol solution for 10s at 25ºC. A 7.5nm thick layer of aluminum was deposited by electron beam evaporation at the rate of 0.1Å/s, followed by liftoff in acetone. The sample was then treated with an oxygen plasma in a low power Diener Plasma Asher for 15s to remove the graphene that was not protected by the aluminum etching mask. After the plasma etching, the samples were soaked in 0.1 molar HCl solution for 2 days to allow complete removal of the aluminum mask, leaving the graphene ribbons. Then 5nm/50nm Cr/Au contacts were patterned and deposited on the ribbons by electron beam lithography and evaporation. Finally, for some samples, an 8nm alumina passivation layer was deposited by electron beam evaporation at the rate of 0.1Å/s. Electrical characterization of each device was performed using a Keithley 4200 Semiconductor Characterization System under vacuum conditions. Experiments were carried out at room temperature (293K) and liquid nitrogen temperature (77K) as well as at an intermediate temperature of 200K.

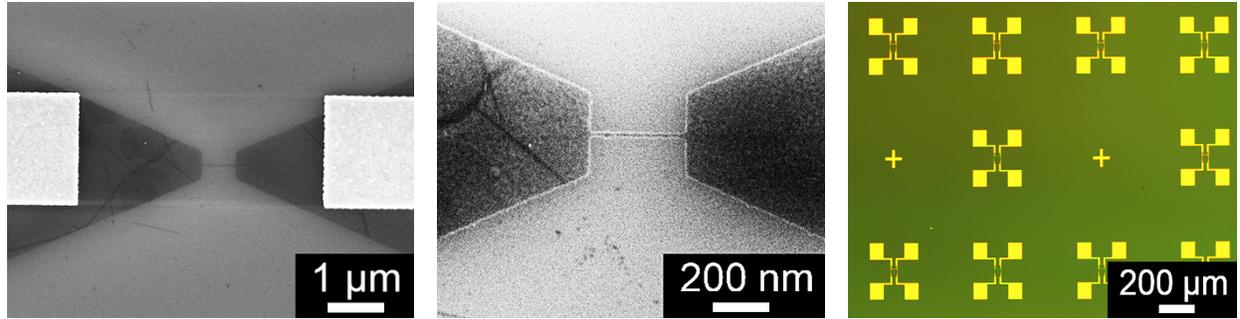

***Figure 2.*** *(a) One of the devices without alumina passivation layer on top. The graphene nanoribbon in the middle is 20nm wide and 600nm long. The Au electrodes are visible on either side; (b) zoom-in on the GNR region; (c) Optical image showing an array of devices arranged on a single chip.*

Figure 2 shows a Scanning Electron Microscope (SEM) image of a 20 nm wide ribbon to illustrate the geometry used. To minimise contact resistance, band-bending and current-crowding effects, the GNR is connected to graphene wedge structures all one piece which are in turn connected to metal electrodes that are several microns away. In order to minimise Joule heating and therefore eliminate any current-induced changes in resistance [27], the current was kept below 10 µA during testing. Subtracting the average resistance of devices that have no GNRs between wedge structures from the resistance of a normal device gives the resistance, *R* of each GNR. A plot of the measured resistivity, i.e. the sheet resistance, $R_s = R\frac{w}{L}$ as a function of the ribbon width is shown in Figure 3(a), from which we can see that the resistivity starts to significantly increase once the width decreases below about 50 nm, in agreement with what others have reported [28, 29] but as our ribbons are smaller than in those studies, we are better placed to test theoretical predictions. In Figure 3(b), we show results for a similar batch of devices, but under vacuum conditions and at three different temperatures between 77K and 293K. This shows the same overall trend, and that the resistivity decreases with increasing temperature, as expected for graphene.

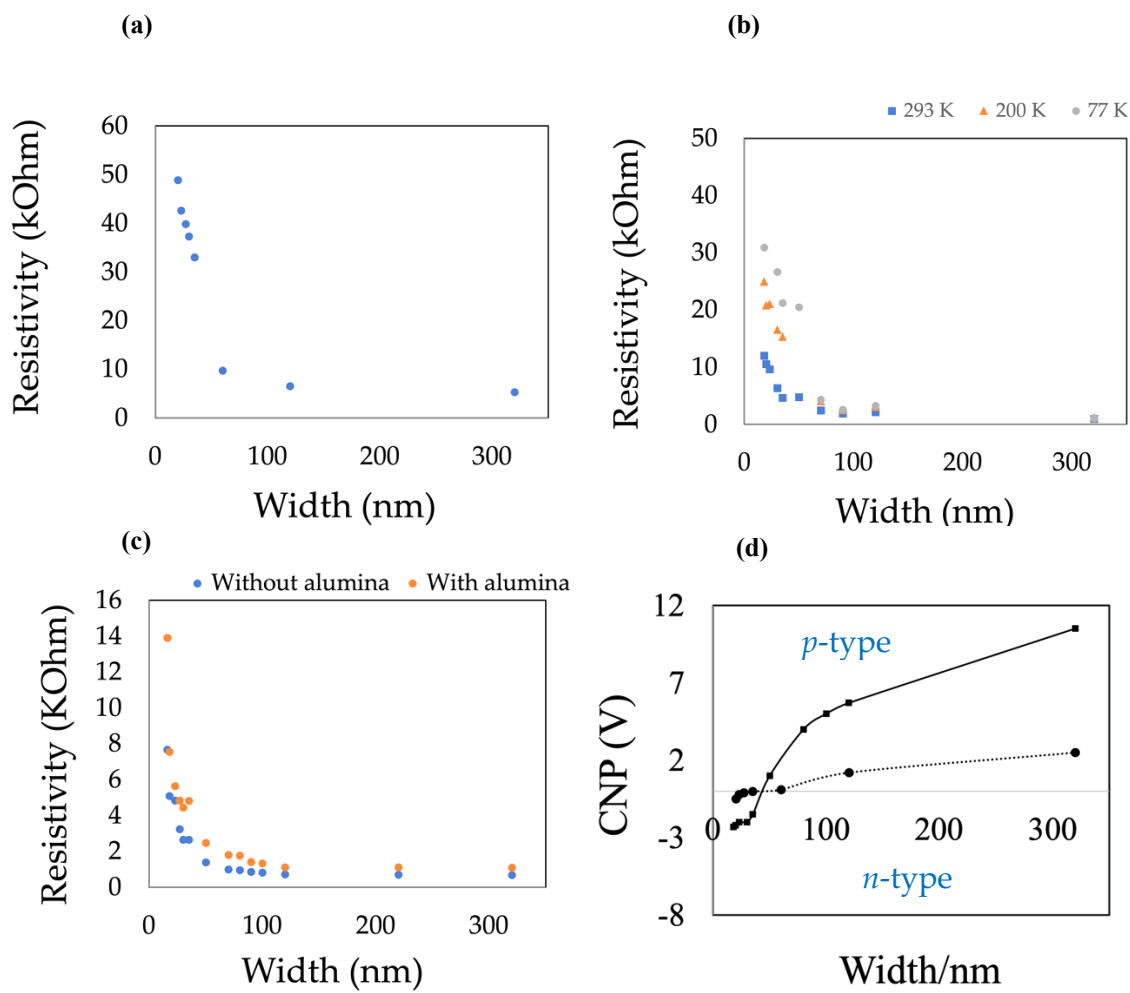

*Figure 3.* *(a) Resistivity Vs width for unpassivated GNRs (600 nm length); (b) Resistivity Vs width for a different batch of unpassivated GNRs as a function of temperature, showing the expected decrease in resistivity with increasing temperature; (c) resistivity Vs width for unpassivated and $Al_2O_3$-passivated devices showing that passivated devices have higher resistivity; (d) Variation of CNP on GNR width shows a transistion in polarity around a width of 50 nm.*

In line with what others have observed, we see that the resistivity varies from one batch to another due to varying levels of contamination arising from the specific fabrication procedures and the age of the graphene sample. It is well known that graphene is particularly susceptible to unintentional doping from the environment, so various strategies have been employed to reduce this including a variety of chemical, thermal and physical treatments. Nonetheless, passivation is a commonly used technique to mitigate against unwanted ageing effects. We have fabricated further devices, but with a top alumina layer, deposited as described above. One would expect that passivated devices will experience less unwanted doping and will therefore have a higher resistivity but will otherwise display the same characteristics as the un-passivated ones. This is verified as shown by the results in Figure 3(c).

In order to understand and explain the dramatic increase in resistivity for ribbon widths below around 50 nm, we ultimately need to explore the nature of conduction in such systems. One route towards this, which is particular to 2D systems, is measurements of the charge neutrality point (CNP), or Dirac point, i.e. the gate voltage at which the conductance of a graphene device reaches its minimum value, as the unintentional dopants are compensated for at this voltage. As shown in Figure 3(d), the CNP also varies with GNR width. We have observed on all devices (multiple batches comprising >100 GNRs) that the CNP switches polarity from predominantly *p*-type to *n*-type once the ribbon width is below around 50 nm. Different batches exhibit different ranges of CNP, with the most contaminated samples having a CNP of order 50V, and the cleanest samples having a CNP close to 0V.

There are several independent effects influencing the resistivity, including edge scattering, dopant concentration and carrier mobility, all of which depend on width. Single-layer graphene is a 2D material, so in principle the charge carriers do not scatter from the top or bottom surface, and instead only see the edges. The effect of the unintentional doping from adsorbed/deposited material on the graphene surface is to create local charge puddles within the graphene which act as local scatterers [30]. Therefore, we can describe the resistivity as comprising two terms: a bulk, width-independent value which is dominated by this doping and the width-dependent term which starts to become relevant for ribbon widths comparable to the carrier mean free path.

Although, as we have stressed, the electronic structure of graphene is fundamentally different to that of a metal or semiconductor, the Fuchs-Sondheimer model is agnostic with respect to any particular conduction mechanism. To calculate the resistivity of a GNR, there are a number of relevant parameters — the mean-free path $\lambda$, the proportion of electrons specularly reflected from the ribbon edges $p$, the ribbon width, $w$ and the bulk resistivity $\rho_0$. The quantity $p$ varies in the range $0 \leq p \leq 1$, where $p = 0$ and $p = 1$ correspond to fully diffuse and fully specular reflection, respectively. The very fact that we experimentally observe an increasing resistivity with decreasing ribbon width indicates that within the framework of this model $p \neq 1$, and the overwhelming experimental evidence in the literature is that $p \sim 0$, i.e. scattering from graphene edges is fully diffuse. From this model, the width-dependent component of resistivity, $\rho_{w(w)}$ for the case of fully diffuse scattering is

$$\rho_w(w) = \rho_0 \left[1 + \frac{3}{8}\frac{\lambda}{w}\right]. \qquad \textit{Equation (1)}$$

Therefore, the only unknown quantity is $\lambda$. Typical values of $\lambda$ in the literature are of the order tens to hundreds of nm at room temperature [12, 14], depending on the specific type of encapsulation used. The longest mean free path reported at room temperature is of the order 500 nm for graphene encapsulated in hexagonal BN [31]. Given that we are instead using an oxide layer which is defect-rich, we expect that 500 nm will be the absolute upper limit on $\lambda$. We can use Equation (1) to predict the expected variation of resistivity on ribbon width for this extreme case, as shown in Figure 4(a). We have taken the uncertainty in ribbon size due to edge roughness and effective width into account by plotting two curves – the upper and lower ones correspond to ribbons 10 nm narrower and 10 nm wider than the nominal width, respectively.

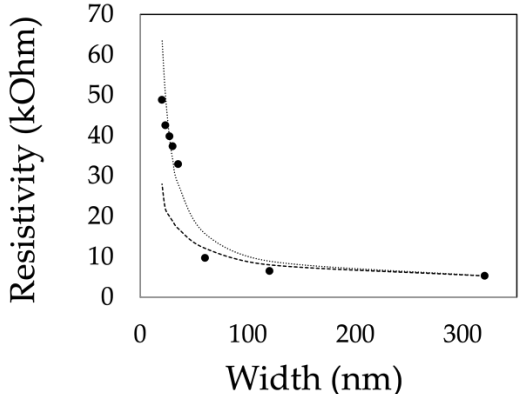

**Figure 4.** Plot of equation (1) for ribbons 10 nm narrower (upper curve) and 10 nm wider (lower curve) than the nominal ribbon width, superimposed on the data from Figure 3(a), for $\lambda = 500$nm.

However, it is not physically reasonable to assume that the mean free path is so long at room temperature or that the wires are 10 nm narrower than their actual physical size. Imaging using SEM and AFM reveals that the physical GNR width fluctuates by ± 2 nm at most. Therefore, this simple model of edge scattering alone is not sufficient to explain our findings that the resistivity has a stronger dependence on width than $w^{-1}$ so a more sophisticated model is needed. It has been shown theoretically that the concept of a unique value of specularity no longer exists for ribbons with edge roughness and with widths below around 50 nm, and that the specific band structure of graphene does need to be considered [32]. Previous studies which looked at wider ribbons would have predicted a better fit to this model so would have significantly over-estimated the mean free path.

We cannot ignore the fact that the CNP and therefore the conductivity depends dramatically on ribbon width, and this must somehow be taken into account. The CNP data presented in Figure 3(d) shows that ribbons narrower than around 50 nm are predominantly *n*-type whereas wider ones are *p*-type. This is simply a consequence of the fact that the edges tend to be disordered, oxidised and have an increased electron density and a higher Fermi level than the bulk (top and bottom) surfaces. We have fabricated GNRs at different orientations and have observed no discernible effect on the measured resistivity, indicating that the edges are highly disordered and neither uniquely armchair nor zigzag. We would expect anomalous transport characteristics at the point where the majority charge carriers switch polarity, which is also where the resistivity starts to change most noticeably. This is further revealed by the temperature dependence of resistance (Figure 3(b)), from which we extract the TCR (Temperature Coefficient of Resistance), which we find to be in the expected range of -0.001 to -0.004 Ohm/degree K [33]. However, a careful analysis of the data reveals that this has a strong dependence on GNR width, as shown in Figure 5, where we see a peak in the TCR at a GNR width of around 40-50 nm, similar to the width at which the CNP switches polarity.

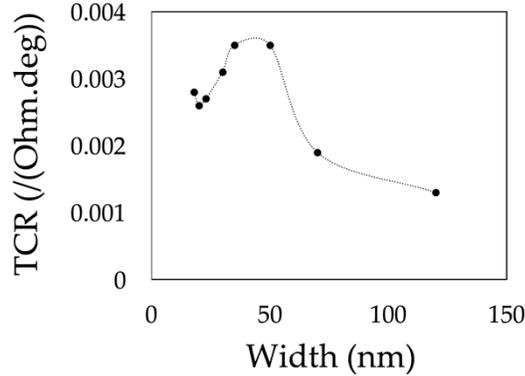

*Figure 5.  Variation of the TCR (Temperature Coefficient of Resistance) vs GNR width showing a peak at around 40 nm, close to the width at which the CNP switches sign*

The peak in TCR indicates that at this width, electron-phonon scattering is enhanced.  The exact mechanism behind this is as yet unclear and warrants further investigation.  This is the width at which the overall GNR has equal amounts of *n*-type and *p*-type doping, and during a transport measurement, equal numbers of electrons and holes will be flowing in opposite directions, increasing the opportunities for carrier scattering.  This is also around the width at which the bandgap starts to become noticeable (several $k_BT$).  Multiple studies have shown that the concentration of charge carriers is increased at the edges of a GNR [17, 18, 30, 34, 35], so it is no surprise that for narrow wires, the edge states can start to dominate.

On the basis of the actual band structure of graphene and within the fully quantum Landauer-Büttiker formalism, we can describe a GNR in terms of two distinct mean free paths, $\lambda_m$ and $\lambda_D$, due to surface (edge) and bulk defect scattering, respectively, where $\lambda_m$ is defined as $\lambda_m = w(v_\parallel/v_\perp)$ and where $v_\parallel$ and $v_\perp$ are the longitudinal and transverse electron velocities of each conduction mode (channel), respectively [28].  This has the form

$$\lambda_m = w\sqrt{\left(\frac{2wE_f}{mhv_f}\right)^2 - 1}.  \qquad \textit{Equation (2)}$$

where $E_f$ and $v_f$ are the Fermi energy and velocity, respectively, typically 0.2 eV and 1x10$^6$ m/s, and *m* refers to the *mth* conduction mode. This is similar to what is predicted from the Fuchs-Sondheimer theory in that it implicitly predicts a width-dependent mean free path due to surface scattering, which we assume is fully diffuse.

The total conductance as described by both scattering mechanisms is

$$G = \frac{2e^2}{h}\sum_m \frac{T_m}{1+L\left(\frac{1}{\lambda_D}+\frac{1}{m}\right)}. \qquad \textit{Equation (3)}$$

Where the summation is over all conduction channels (of energy $E_m$), i.e. for $E_m \leq E_f$ and each channel has a quantum transmission probability $T_m$. The number of channels is given by the number of electron modes that can fit across the GNR width, which is $2w/\lambda_F$ where $\lambda_F$ is the Fermi wavelength of the electrons. From this expression and assuming that all channels have the same transmission probability, $T_m = 1$, we calculate the effective resistivity to be

$$\rho = \frac{h}{2e^2}\left(\frac{1}{\frac{\lambda_D}{w}+1.5w^{1.5}\sqrt{\left(\frac{2E_f}{hv_f}\right)^3}}\right). \qquad \textit{Equation (4)}$$

This expression predicts that resistivity varies mostly as $w^{-\frac{3}{2}}$, as opposed to the prediction of $w^{-1}$ based on surface scattering alone that we saw earlier. The one issue we must still address however, is that of the variation we observe in CNP Vs width. The CNP is a measure of the charge density, which is related to the Fermi level, $E_f$, so to first order we can take the assumption that $E_f$ is of the form $E_f = B(w_0-w)$, where $B$ is a constant, to be determined individually for each batch of devices, and $w_0$ is the width at which the CNP switches polarity. This form takes into account the fact that our ribbons become increasingly *n*-type (corresponding to a higher $E_f$) as width decreases.

In reality, the sign of $E_f$ is not relevant for our evaluation of the mean free path, so we will only consider its absolute value. Combining this empirical finding with the calculation above, we obtain the following expression for resistivity of a GNR:

$$\rho = \frac{h}{2e^2}\left(\frac{1}{\frac{\lambda_D}{w}+1.5w^{1.5}\sqrt{\left(\frac{2B|w_0-w|}{hv_f}\right)^3}}\right).\qquad \textit{Equation (5)}$$

In Figure 4, we show the fits of this formula to the resistivity Vs width data with one fitting parameter: bulk mean free path ($\lambda_D$). The fits are sufficiently good that we can be confident they verify the above model. The values for $\lambda_D$ that we obtain are in Table 1. Using this model, we find that the mean free path is up to 220 nm for unpassivated devices and down to 72 nm for Alumina-passivated ones, in agreement with expectations [4, 5, 14]. Having validated the model summarized by Equation (5), with a few simple measurements of parameters ($\lambda_D$, $B$, $w_0$) specific to a given process is possible to predict the resistivity of any GNR, which paves the way towards designer interconnects and devices.

Table 1. Device parameters for the quantum model (equation (5)).

| Device Number | Al Passivation layer | Mean free path, $\lambda_D$ (nm) |
|---|---|---|
| 1 | No | 150 |
| 2 | No | 220 |
| 3 | No | 105 |
| 4 | Yes | 72 |

To summarise, we have shown that graphene nanoribbons with width below around 100 nm display some unusual characteristics not seen in larger structures, namely significant edge scattering that can be described as fully diffuse, a CNP that switches sign at a GNR width of around 50 nm, and evidence for a peak in the electron-phonon scattering rate at or around the same width. The addition of an alumina passivation layer may reduce sensitivity to atmospheric conditions but it also significantly reduces the mean free path for conduction. We have demonstrated that the width-dependence can be explained by a quantum model which assumes fully diffuse edge scattering and which allows us to extract the effective mean free path for conduction.

## ACKNOWLEDGEMENTS

The authors would like to thank Juan Rubio-Lara for advice on fabricating the graphene nano-ribbons.

## Author Contributions & Competing interests

CD devised the experiment, performed the calculations and modelling, supervised the work and contributed to writing the article. ZX carried out the experimental work and contributed to writing the article. There are no competing interests.